\documentstyle[preprint,aps]{revtex}

\begin{document}
\draft

\title{Nonexistence of a $\eta NN$ quasibound state}

\author{H. Garcilazo $^{(1)}$
and M. T. Pe\~na $^{(2),(3)}$}

\address{$(1)$ Escuela Superior de F\' \i sica y Matem\'aticas \\
Instituto Polit\'ecnico Nacional, Edificio 9,
07738 M\'exico D.F., Mexico}

\address{$(2)$ Centro de F\'\i sica das Interac\c c\~oes Fundamentais, \\
Av. Rovisco Pais, P-1049-001 Lisboa, Portugal }

\address{$(3)$ Departamento de F\'\i sica, Instituto Superior T\'ecnico,\\
Av. Rovisco Pais, P-1049-001 Lisboa, Portugal}

\maketitle

\begin{abstract}
We have solved the Faddeev equations for 
$\eta d$ elastic scattering 
using realistic separable interactions for the $NN$ and coupled 
$\eta N$-$\pi N$ subsystems.
We found that including explicitly
the pion channel in the integral equations  
drastically reduces the attraction that is present in the system.
As a consequence, the existence of a $\eta NN$ quasibound state is excluded
by the modern $\eta N$ amplitude analysis. 

\end{abstract}

\pacs{14.40.Ag,25.40.Ve,25.80.Hp}

\narrowtext
\newpage

The possible existence of a $\eta NN$ quasibound state in the 
$\eta d$ system was first suggested by Ueda \cite{UEDA}. He solved
the Faddeev equations of elastic $\eta d$ scattering using separable 
interactions for the $NN$ and coupled $\eta N$-$\pi N$ subsystems. At the
time of Ueda's prediction, however, very little was known about the 
$\eta N$ channel, so that he
fitted his $\eta N$ and $\pi N$ interactions basically to the $\pi N$
data only. He found that his model predicted the existence of a $\eta NN$
quasibound state very near threshold with a mass of 2430 MeV and a 
width of 10-20 MeV.

More recent calculations \cite{RAKI,GRE1,SHEV,DELO} have used in the case of 
the coupled $\eta N$-$\pi N$ subsystem only the $\eta N$ subsector by 
means of a Yamaguchi separable potential with a complex 
energy-dependent strength. They found that the existence of the 
quasibound state depended strongly on the value of the real part 
of the $\eta N$ scattering length, such that $Re$ $a_{\eta N}\approx 1$
fm is required in order for the quasibound state to exist.
This value of $Re$ $a_{\eta N}$ is within the range of values given by
modern $\eta N$ amplitude analysis \cite{BAT1,GRE2,GRE3}.

However, in a very recent paper \cite{GAR1} we have pointed out that a true 
measure 
of the attraction or repulsion present in a three-body system can only be
obtained by assuming two-body interactions which are real and energy
independent. Therefore, in Ref. \cite{GAR1}
we constructed separable potential models 
of the coupled $\eta N$-$\pi N$ subsystem in which the strength of the
potentials is real and energy independent, so that the imaginary part
of the $\eta N$ scattering length is generated by the coupling to
the $\pi N$ channel. These models were required to fit not only the
$\eta N$ scattering length but also the $\eta N$ amplitude in the 
vicinity of the $S_{11}$ resonance as obtained by the recent $\eta N$
data analysis \cite{BAT1,GRE2,GRE3}. In Ref. \cite{GAR1} we
used the diagonal $\eta N\to\eta N$ part of the full $\eta N$-$\pi N$
amplitude to calculate $\eta d$ elastic scattering in a truncated
approximation where the pion was not included explicitly in the 
integral equations but only implicitly through his contribution
in the propagator of the $\eta N$ interacting pair. We used for 
the $NN$ interaction in the $^3S_1$ channel the so-called PEST separable
potential \cite{PEST} which takes into account the $NN$ repulsion
at short distances. We found in Ref. \cite{GAR1} that the truncated
model does not give rise to a $\eta NN$ quasibound state for any
of the models based on modern $\eta N$ amplitude analyses. However,
two question that immediately arise are a) how important is the 
explicit contribution of the pion? and b) is it attractive or
repulsive? We will answer these two questions in this paper.

In Ref. \cite{GAR1} we constructed 6 different phenomenological
models of the coupled $\eta N$-$\pi N$ subsystem which were fitted
to the  amplitudes of recent data analyses
\cite{BAT1,GRE2,GRE3}. All the 6 potentials have separable form

\begin{equation}
<p|V_{\eta\eta}|p'>=-g_\eta(p)g_\eta(p'),
\label{eq24}
\end{equation}

\begin{equation}
<p|V_{\pi\pi}|p'>=-g_\pi(p)g_\pi(p'),
\label{eq26}
\end{equation}

\begin{equation}
<p|V_{\eta\pi}|p'>=\pm g_\eta(p)g_\pi(p'),
\label{eq25}
\end{equation}
with

\begin{equation}
g_\eta(p)=\sqrt{\lambda_\eta}\,{A+p^2\over (\alpha_2^2+p^2)^2},
\label{eq46p}
\end{equation}

\begin{equation}
g_\pi(p)=\sqrt{\lambda_\pi}\,{1 \over \alpha_\pi^2 + p^2}.
\label{eq37}
\end{equation}

The parameters of the six models are given in table III of Ref. 
\cite{GAR1}. If one substitutes the potentials (\ref{eq24})-(\ref{eq25})
into the Lippmann-Schwinger equation of the coupled $\eta N$-$\pi N$
subsystem one obtains that
the T-matrices are of the form

\begin{equation}
<p|t_{\eta\eta}(E)|p'>=g_\eta(p)\tau_2(E)g_\eta(p'),
\label{eq28}
\end{equation}

\begin{equation}
<p|t_{\pi\pi}(E)|p'>=
g_\pi(p)\tau_2(E)g_\pi(p'),
\label{eq30}
\end{equation}

\begin{equation}
<p|t_{\eta\pi}(E)|p'>=\pm
g_\eta(p)\tau_2(E)g_\pi(p'),
\label{eq29}
\end{equation}
where

\begin{equation}
\tau_2(E)=[-1-G_\eta(E)-
G_\pi(E)]^{-1},
\label{eq31}
\end{equation}

\begin{equation}
G_\eta(E)=\int_0^\infty p^2 dp{g_\eta^2(p) \over E-p^2/2\mu_2+i\epsilon},
\label{eq32}
\end{equation}

\begin{equation}
G_\pi(E)=\int_0^\infty p^2 dp{g_\pi^2(p) \over E +p_0^2/2\mu_\pi
-p^2/2\mu_\pi+i\epsilon}.
\label{eq33}
\end{equation}
$\mu_2$ and $\mu_\pi$ are the $\eta N$ and $\pi N$ reduced masses 
respectively while $p_0$ is the $\pi N$ relative momentum at the
$\eta N$ threshold, i.e., 

\begin{equation}
p_0^2={[s_0-(m_\pi+m_N)^2][s_0-(m_\pi-m_N)^2] \over 4s_0},
\label{eq34}
\end{equation}
with

\begin{equation}
s_0=(m_\eta+m_N)^2.
\label{eq35}
\end{equation}

Thus, the Faddeev equations for $\eta d$ elastic scattering take the form 
diagrammatically depicted in Fig. 1.
In the second equation of this figure, there is only a term
with a nucleon-nucleon interaction proceeding while a meson (the $\eta$)
is a spectator, since the term where
the spectator meson is a pion involves an intermediate state (formed by
a pion and a $NN$ state in the $^3S_1$ channel) of isospin 1, 
while the $\eta d$ system has isospin 0. Similarly, the intermediate 
state where a pion is the spectator and the $NN$ state is in the $^1S_0$
channel can not proceed either due to the fact that this 
state has total spin 0, while the $\eta d$ system has total spin 1.

The integral equation depicted in Fig. 1 has the analytical form

\begin{equation}
T_2(q_2;E) = 2K_{21}(q_2,q_{10};E) + \int_0^\infty {q_2^\prime}^2 dq_2^\prime
M(q_2,q_2^\prime;E)\tau_2(E-{q_2^\prime}^2/2\nu_2)T_2(q_2^\prime;E),
\label{eq2}
\end{equation}
where the kernel $M(q_2,q_2^\prime;E)$ is given by

\begin{eqnarray}
M(q_2,q_2^\prime;E) & = & K_{23}(q_2,q_2^\prime;E) - 
K_{23}^\pi(q_2,q_2^\prime;E) \nonumber \\ 
& & + 2\int_0^\infty q_1^2 dq_1
K_{21}(q_2,q_1;E)\tau_1(E-q_1^2/2\nu_1)K_{12}(q_1,q_2^\prime;E).
\label{eq3}
\end{eqnarray}
If one drops the term
$K_{23}^\pi$, Eqs. (\ref{eq2}) and (\ref{eq3}) are identical to Eqs.
(2) and (3) of Ref. \cite{GAR1}. The kernels $K_{ij}$ have been defined in
\cite{GAR1} and the new term 
$K_{23}^\pi$ is equal to $K_{23}$ except that particle 1 is now a 
$\pi$ instead of a $\eta$. 

We note at this point that the $\eta$N $\rightarrow$ $\pi$N transition amplitude,
describing a pion exchange followed by an $\eta$ exchange, enters
an even number of times at every order of iteration of the integral
equation in Fig. 1 (i.e.
equation (15)). Therefore, the ambiguity in sign in the $\eta$N $\rightarrow$ $\pi$N transition amplitude, 
explicit in equations (3) and (8),
is immaterial for this calculation. 

The most important point in Eq. (\ref{eq3}) is that $K_{23}$ and
$K_{23}^\pi$ appear with opposite signs. These signs come from the 
reduction of the Faddeev equations when one has two identical fermions
\cite{AFNA,GAR2}. Since we are assuming that the meson is particle 1
so that 2 and 3 are the two fermions and all orbital angular momenta are
equal to zero, then following the reduction procedure of Refs.
\cite{AFNA,GAR2} leads to the result that the kernel $K_{23}$ must
by multiplied by a factor $F_{23}$, where

\begin{equation}
F_{23}=F_{23}^{Identical}F_{23}^{Spin}F_{23}^{Isospin},
\label{eq4}
\end{equation}
and

\begin{equation}
F_{23}^{Identical}=-(-)^{s_1+s_3-S_2+i_1+i_3-I_2},
\label{eq5}
\end{equation}

\begin{equation}
F_{23}^{Spin}=(-)^{S_3+s_3-S}\,\sqrt{(2S_2+1)(2S_3+1)}\,W(s_3s_1Ss_2;S_2S_3),
\label{eq6}
\end{equation}

\begin{equation}
F_{23}^{Isospin}=(-)^{I_3+i_3-I}\,\sqrt{(2I_2+1)(2I_3+1)}\,
W(i_3i_1Ii_2;I_2I_3),
\label{eq7}
\end{equation}
with $W$ the Racah coefficient, and $s_i$, $S_i$, and $S$ ($i_i$, $I_i$, 
and $I$) are the spins (isospins)
of particle i, of the 
pair jk, and the three-body system.
It is straightforward to see that the factor $F_{23}$ is equal to 1
when particle 1 is a $\eta$ but it is equal to -1 when particle 1 is
a $\pi$. All other spin-isospin recoupling coefficients that would appear 
in Eqs. (\ref{eq2}) and (\ref{eq3}) are equal to 1.

In Eq. (\ref{eq2}) the propagator 
$\tau_2(E-{q_2^\prime}^2/2\nu_2)$ is the one appropriate for a $\eta N$
interacting pair since $\nu_2$ is the reduced mass of a nucleon and a 
$\eta N$ pair. In principle, one should have two $\eta$-N amplitudes 
corresponding to the two possibilities of decay for the $S_{11}$ isobar, into
a $\eta N$ or a $\pi N$ pair. However, if one assumes that

\begin{equation}
\tau_2(E-{q_2^\prime}^2/2\nu_2)=
\tau_2(E-{q_2^\prime}^2/2\nu_\pi),
\label{eq8}
\end{equation}
where $\nu_\pi$ is the reduced mass of a nucleon and a $\pi N$ pair,
one obtains a single equation. We have checked numerically that 
the effect of separating Eq. (\ref{eq2}) into two equations,  that is 
of considering

\begin{equation}
\tau_2(E-{q_2^\prime}^2/2\nu_2)\ne
\tau_2(E-{q_2^\prime}^2/2\nu_\pi),
\label{eq9}
\end{equation}
is to produce changes in the $\eta d$ scattering length of less that 1 \%. 
We should point out that in a relativistic Faddeev theory \cite{GAR3}
the energy of the isobar is independent of the mode into which it
decays so that the equivalent of Eq. (\ref{eq8}) always holds.

We solved the integral equation (\ref{eq2}) using
the method of contour rotation \cite{HETH}. We give in table I the
results for the $\eta d$ scattering length for the case of the impulse
approximation and for the full calculation with and without the
pion contribution. As one sees, the effect of including the pion
channel explicitly is quite large and it reduces the $\eta d$ 
scattering length. This reduction is a direct 
consequence of the minus sign in front
of the kernel $K_{23}^\pi$ representing the pion contribution. 
The equations for $\eta d$ elastic scattering 
without the pion contribution were not attractive enough 
to produce a $\eta NN$ quasibound state
(the signal that a quasibound state exists for a given model is that the
real part of $A_{\eta d}$ becomes negative while the imaginary part 
becomes large), but it turns out that the inclusion of the pion
reduces even further the attraction,
completely ruling out the existence of a quasibound state in this 
system. It is worth pointing out that the minus sign for the second
term of the right-hand-side of equation (15) is critical: 
if one takes the pion
contribution with a plus sign instead of the correct minus sign, the six 
models of the coupled $\eta N$-$\pi N$ subsystem will give rise to
a quasibound state in the $\eta d$ system. 

We show in Fig. 2 the results for the cross section of $\eta d$ elastic
scattering in the region near threshold again for the cases of the 
impulse approximation and the full calculation with and without the
pion contribution. As one sees, the strong enhancement of the cross 
section near threshold is greatly reduced when the pion contribution
is included. Unexpectedly, one re-encouters here the pattern of cancellation between the $\pi$ and $\eta$ re-scattering processes found in reference \cite{PIDETANN},
in a one-loop calculation for 
the $\pi$d $\rightarrow$ $\eta$NN reaction.

To conclude, we have shown that the explicit contribution of the pion
drastically reduces the amount of attraction that is present in the
$\eta d$ system, such that there is no possibility for a $\eta NN$
quasibound state to exist in this system.

This work was supported in part by 
COFAA-IPN (M\'exico) and
by Funda\c c\~ao para a Ci\^encia e a Tecnologia,
MCT, under contracts PRAXIS XXI/BCC/18975/98 and
PRAXIS/P/FIS/10031/1998 (Portugal).

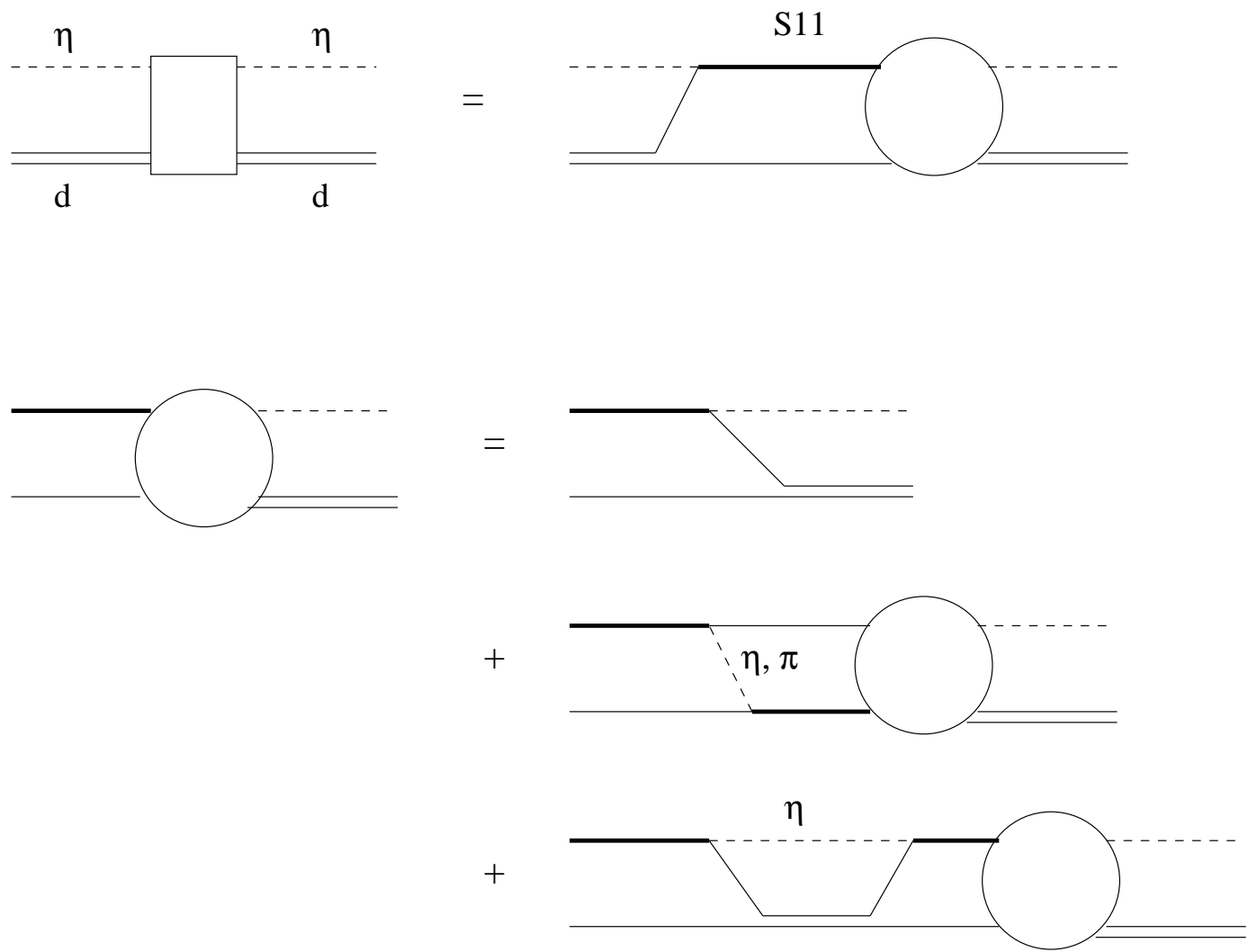
\begin{figure}
\caption{Faddeev equations for $\eta d$ elastic scattering.} 
\end{figure}

\begin{figure}
\caption{Integrated $\eta d$ elastic cross sections of the three-body model
with the pion contribution (solid lines), of the
three-body model without the pion contribution (dashed lines), 
and of the impulse approximation (dot-dashed lines)
for the six models of the $\eta N$-$\pi N$ subsystem,
as a function of the c.m. $\eta d$ kinetic
energy.}
\end{figure}

\begin{table}
\caption{$\eta d$ scattering length (in fm) for the six models of the 
$\eta N$-$\pi N$ subsystem.
The first column indicates
the reference of the $\eta N$-$\pi N$ amplitude analysis 
on which the model is based, the
second column indicates the $\eta N$ scattering length (in fm) of that model,
the third column gives $A_{\eta d}$ from the impulse approximation,  
the fourth column gives $A_{\eta d}$ from the full model without pion   
contribution, and the fifth column gives $A_{\eta d}$ from the full model
with pion contribution.}  

\begin{tabular}{ccccccc}
& Ref. & $a_{\eta N}$ &  Impulse  & Full ($\eta$)  
& Full ($\eta+\pi$)  &  \\
\tableline
& [6]    & 0.72+i0.26 & 1.33+i0.36  & 2.46+i1.62 & 1.55+i0.49 & \\
& [7]    & 0.75+i0.27 & 1.37+i0.36  & 2.61+i1.72 & 1.65+i0.53 & \\
& [8](D) & 0.83+i0.27 & 1.48+i0.34  & 3.10+i2.03 & 1.96+i0.62 & \\
& [8](A) & 0.87+i0.27 & 1.52+i0.34  & 3.36+i2.19 & 2.12+i0.67 & \\
& [8](B) & 1.05+i0.27 & 1.74+i0.30  & 4.81+i3.19 & 3.03+i0.96 & \\
& [8](C) & 1.07+i0.26 & 1.76+i0.29  & 5.02+i3.14 & 3.17+i0.98 & \\
\end{tabular}
\end{table}

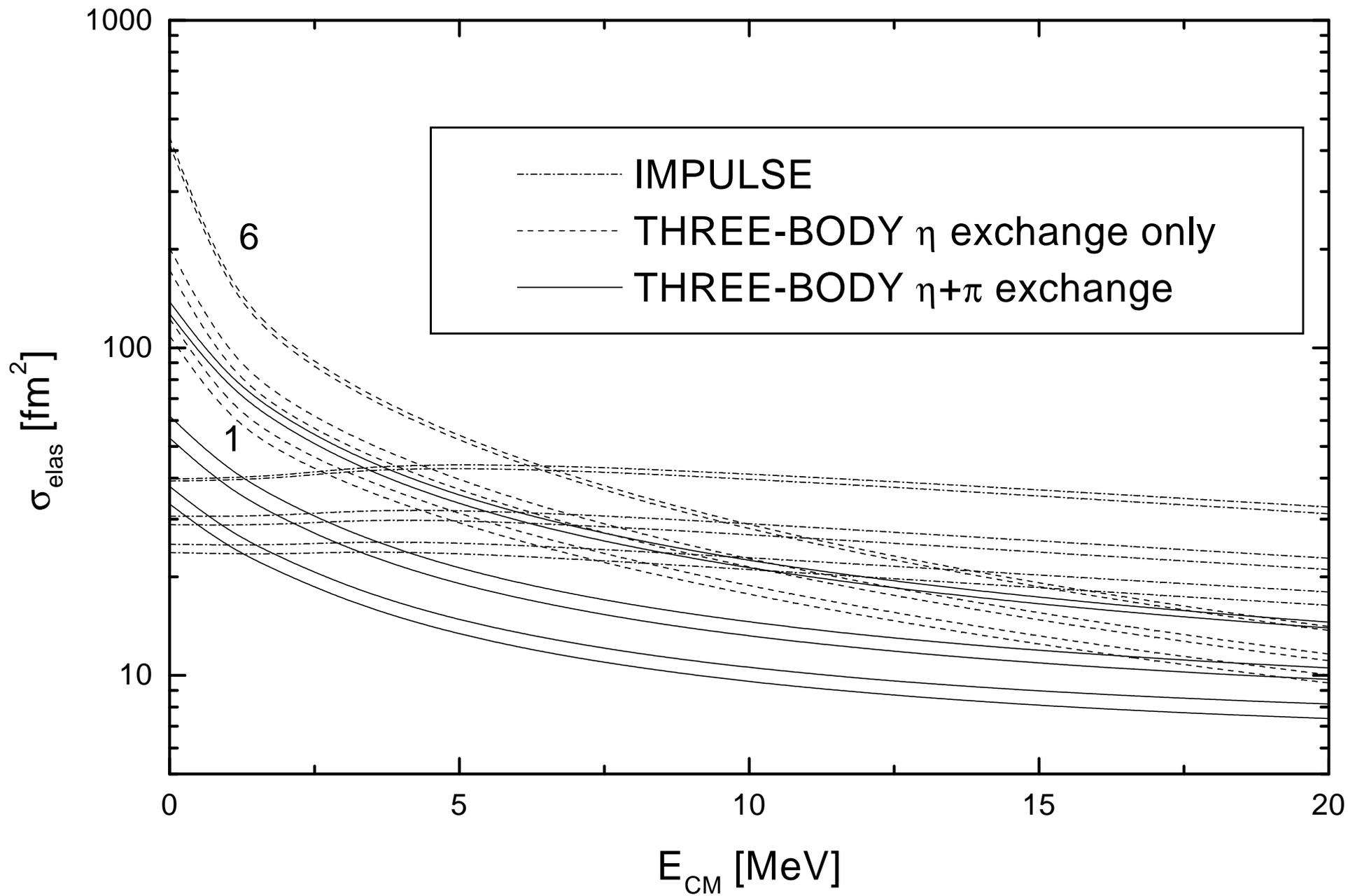
\end{document}